\documentclass[aps,prl,showpacs,twocolumn]{revtex4-1}

\usepackage{graphicx}
\usepackage{bm}
\usepackage{amssymb} 
\usepackage{amsmath} 
\usepackage{amsthm}
\usepackage{hyperref}
\usepackage[utf8]{inputenc}
\hypersetup{
     colorlinks=true,      
    linkcolor=blue,        
    citecolor=blue,
    filecolor=magenta, 
    urlcolor=black          
}

\providecommand{\abs}[1]{\left\lvert#1\right\rvert}

\providecommand{\bra}[1]{\langle #1 \rvert}
\providecommand{\ket}[1]{\lvert #1 \rangle}

\providecommand{\ketbra}[2]{\lvert  #1\rangle \langle #2 \rvert}
\newcommand{\expec}[1]{\langle #1\rangle}

\usepackage{mathbbol}

 \newtheorem{definition}{Definition}
 \newtheorem{theorem}{Theorem}

\begin{document}
\title{Characterizing multipartite entanglement with moments of random correlations}

\author{Andreas Ketterer}
\author{Nikolai Wyderka}
\author{Otfried G\"uhne}
\affiliation{Naturwissenschaftlich-Technische Fakult\"at, Universit\"at Siegen, Walter-Flex-Str. 3, 57068 Siegen, Germany}

\begin{abstract}
The experimental detection of multipartite entanglement usually requires a number of appropriately chosen local quantum measurements which are aligned with respect to a previously shared common reference frame. The latter, however, can be a challenging prerequisite e.g. for satellite-based photonic quantum communication, making the development of alternative detection strategies desirable. One possibility for avoiding the distribution of classical reference frames is to perform a number of local measurements with settings distributed uniformly at random. In this work we follow such a treatment and show that an improved detection and characterization of multipartite entanglement is possible by combining statistical moments of different order. To do so, we make use of  designs which are pseudo-random processes allowing to link the present entanglement criteria to ordinary reference frame independent ones. The strengths of our methods are illustrated in various cases starting with two qubits and followed by more involved multipartite scenarios. 
\end{abstract}

\maketitle 

\textit{Introduction.}--- The key role of multipartite entanglement as a resource in quantum information theory manifests itself through a variety of applications which gave rise to a growing commercial interest in quantum technologies \cite{HorodeckiReview,NielsenChuang,flagship}. Prominent examples are quantum computation or communication protocols which have been shown to outperform known classical counterparts \cite{NielsenChuang,TelepReview,CryptoReview}. Nevertheless, its experimental detection and characterization still comes along with major technical and conceptual difficulties. One such difficulty is the alignment of  local measurement settings among the involved spatially separated parties which is a prerequesite for the evaluation of many entanglement criteria or Bell inequalities \cite{OtfriedReview,BELL,ReviewNonlocality}.

Several proposals allowing to circumvent the problem of measurement alignment have been made. One is to restrict measurements to the single particle level which has proven useful for the characterization of multipartite entanglement under the assumption that the state is pure \cite{GrossPolytopes,StevePolytopes}. Other possibilities are to encode logical qubits into rotational invariant subspaces of combined degrees of freedom \cite{LeandroRotInv,SteveAlignmentFreeComm}, or to exploit the local-unitary (LU) invariance of entanglement criteria based on correlation functions \cite{BriegelLUinv,JulioMarkus,KlocklHuber,JulioCorrMat1,JulioCorrMat2,ZukowskiRefFrame1,ZukowskiRefFrame2,LawsonRefFrame}. The latter, commonly referred to as reference-frame (RF) independent entanglement criteria, can be evaluated without aligning spatially separated measurements \cite{ReviewRefFrames}, but still require the experimenters' control over the choice of local measurement bases, e.g. three orthogonal ones.

Other strategies for RF independent entanglement detection   lift also the last assumption in the sense that only measurements with randomly chosen settings are required. In this case, one has to resort to statistical tools which allow to infer the entanglement properties of the considered states. For instance, in Refs.~\cite{tran1,tran2,LukRud} the authors study entanglement detection given distributions of correlation functions obtained from local measurements with settings chosen uniformly at random. Similarly, one can probe the violation of Bell inequalities with randomly distributed measurement settings \cite{RudolphRandMeasBell,BrunnerRandMeasBell,DiamantiRandMeasBell}.

The latter attempts motivate us to push forward in this direction and show how to considerably improve entanglement criteria based on randomly measured correlation functions \cite{tran1,tran2}. In this work we thus demonstrate that a better entanglement detection and even a characterization of different classes of multipartite entanglement is possible by combining statistical moments of different order. In this respect, we will see that every such statistical entanglement criterion can be traced back to a RF independent one using pseudo-random processes, also referred to as designs. Further on, we demonstrate their strengths for the detection and characterization of multipartite entanglement involving the two lowest non-vanishing moments. We start with the instructive bipartite case of two qubits and then move to the more involved multipartite scenarios. 

\begin{figure}[t]
\begin{center}
\includegraphics[width=0.47\textwidth]{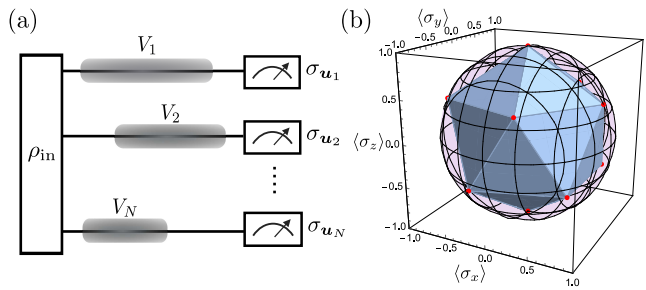}%
\end{center}
\caption{(a) Schematic representation of a $N$-qubit entanglement detection scheme based on local measurements with randomly chosen settings $\boldsymbol u_1,\ldots,\boldsymbol u_N$. The moments $\mathcal R^{(t)}$ are invariant under LU transformations $\{V_n\}_{n=1,\ldots,N}$ (grey shaded areas), which might be caused by the qubits propagation, e.g. the propagation of optical qubits through fiber. (b) Plot of the $6$ measurement settings corresponding to the spherical $5$-design. Each direction yields two points on the Bloch sphere which together yield the vertices of an icosahedron, i.e. a polyhedron with twenty equilateral triangular faces.}
\label{fig_1}
\end{figure}

\textit{Moments of random correlations.}--- To set the stage, let us consider $N$ qubits prepared in the initial state $\rho_\text{in}$ 
which are measured locally according to  the random bases $\{(\ket{u_n^{(0)}}= U_n\ket{0_n},\ket{u_n^{(1)}}= U_n\ket{1_n})\}_{n=1,\ldots,N}$, where the $\{U_n\}_{n=1,\ldots,N}$ represent random unitary transformation picked from the unitary group $\mathcal U(2)$, e.g. according to the Haar measure. We  associate to the random basis $(\ket{u_n^{(0)}}= U_n\ket{0_n},\ket{u_n^{(1)}}= U_n\ket{1_n})$ of the $n$-th qubit a direction $\boldsymbol u_n$ on the Bloch sphere, defined by the components $[\boldsymbol u_n]_i=\mathrm{tr}[\sigma_{\boldsymbol u_n}\sigma_i]$, where $\sigma_i$, with $i=x,y,z$, denote the usual Pauli matrices and $\sigma_{\boldsymbol u_n}= U_n \sigma_z U_n^\dagger$ (see Fig.~\ref{fig_1}(a)). One choice of such set of local random measurement bases leads to the (random) correlation function:
\begin{align}
E(\boldsymbol u_1,\ldots,\boldsymbol u_N)=\expec{ \sigma_{\boldsymbol u_1}\otimes \ldots \otimes \sigma_{\boldsymbol u_N}}_{\rho_\text{in}}.
\label{eq:CorrelationFct}
\end{align}

However, as the directions $\boldsymbol u_n$ are chosen randomly, only one set of random measurement settings will not give any insight  into  the nonlocal properties of the initial state $\rho_{in}$. To achieve this we have to perform several rounds of random measurements and seek a statistical treatment of the obtained values of the correlation function (\ref{eq:CorrelationFct}) through its moments. In order to predict the outcome of this approach we assume that the local measurement directions $\{\boldsymbol u_n\}_{n=1,\ldots,N}$ are chosen uniformly from the Bloch sphere which corresponds to the choice of Haar random unitaries $\{U_n\}_{n=1,\ldots,N}$. 
In this scenario the corresponding moments read:
\begin{align}
\mathcal R^{(t)}&= \frac{1}{(4\pi)^N} \int_{S^{2}} d\boldsymbol u_1\ldots \int_{S^{2}} d\boldsymbol u_N  E(\boldsymbol u_1,\ldots,\boldsymbol u_N)^t, 
\label{eq:RandomMoments}
\end{align}
where $t$ is a positive integer and $d\boldsymbol u_i=\sin{\theta_i} d\theta_i d\phi_i$ denotes the uniform measure on the Bloch sphere $S^{2}$. As the integrals in Eq.~(\ref{eq:RandomMoments}) can be rewritten in terms of integrals with respect to Haar measures on $\mathcal U(2)$, the moments $\mathcal R^{(t)}$ are by definition LU invariant and thus good candidates for RF independent entanglement detection. 
Also, due to the symmetry of the correlation functions (\ref{eq:CorrelationFct}) we can already conclude that $\mathcal R^{(t)}=0$, for all odd $t$ (see App.~A).

\textit{Moments from designs.}--- In order to evaluate the uniform averages over the Bloch sphere in Eq.~(\ref{eq:RandomMoments}) we can resort to so-called spherical $t$-designs which consist of a finite set of points $\{\boldsymbol u_k\in  S^2|k=1,\ldots,L^{(t)}\}\subset S^{2}$  fulfilling the property 
\begin{align}
\frac{1}{L^{(t)}} \sum_{k=1}^{L^{(t)}} P_{t}(\boldsymbol u_k) =\int_{S^{2}}d\boldsymbol u \ P_{t}(\boldsymbol u),
\label{eq:tDesignDef}
\end{align} 
for all homogeneous polynomials $P_{t}:S^2\rightarrow \mathbb R$ of degree at most $t$ \cite{existence}. As $E(\boldsymbol u_1,\ldots,\boldsymbol u_N)^t$ is such a polynomial in each of its local settings $\boldsymbol u_k$, Eq.~(\ref{eq:tDesignDef}) directly yields the formula:
\begin{align}
\mathcal R^{(t)}=\frac{1}{(L^{(t)})^N} \sum_{k_1,\ldots,k_N=1}^{L^{(t)}} E(\boldsymbol u_{k_1}, \ldots  \boldsymbol u_{k_N})^t,
\label{eq:RandomMomentsDesign}
\end{align}
where $\{\boldsymbol u_{k_j}|k_j=1,\ldots,L^{(t)}\}$, for all $j$, are spherical $t$-designs. Hence, we find as a first result that spherical $t$-designs allow for an evaluation of the $\mathcal R^{(t)}$'s  based on a finite number $L^{(t)}$ of local measurement settings and thus directly link them to RF independent entanglement criteria \cite{BriegelLUinv,JulioMarkus,KlocklHuber,JulioCorrMat1,JulioCorrMat2,ZukowskiRefFrame1,ZukowskiRefFrame2,LawsonRefFrame}. 
Furthermore, we note that similar implications also hold for systems of larger local dimensions where one has to resort to unitary designs for the evaluation of the respective moments (see App.~B and C for more details). 

The drawback of spherical and unitary designs is that, while their existence has been proven \cite{existence}, there is no general strategy known to construct them for a given $t$. Nonetheless, by exploiting group theoretical methods it was possible to find a number of examples of exact spherical \cite{ExamplesSphericalDesigns} and unitary designs \cite{Gross5Design}. For instance, a well-known example is the Clifford group which forms a unitary $3$-design and for a qubit reduces to a spherical $3$-design on the Bloch sphere consisting of $L^{(3)}=6$ orthogonal directions $\{\pm\boldsymbol e_i|i=x,y,z\}$.  Furthermore, a number of finite rotation groups on the Bloch sphere were identified as spherical designs of order $t\leq20$ \cite{ExamplesSphericalDesigns}. An example of such a spherical design, with $t=5$, is given by the vertices $\{\boldsymbol v_i|i=1,\ldots,L^{(5)}=12\}$ forming the polyhedron presented in Fig.~\ref{fig_1}(b). Hence, following Eq.~(\ref{eq:RandomMomentsDesign}), we find
\begin{align}
\mathcal R^{(2)}&=\frac{1}{3^N} \sum_{i_1,\ldots,i_N=x,y,z} E(\boldsymbol e_{i_1}, \ldots  \boldsymbol e_{i_N})^2,\label{eq:RandomMoment2}  \\
\mathcal R^{(4)}&=\frac{1}{6^N} \sum_{i_1,\ldots,i_N=1}^{6} E(\boldsymbol v_{i_1}, \ldots  \boldsymbol v_{i_N})^4.
\label{eq:RandomMoment4}
\end{align}
where the limits, $L^{(3)}/2=3$ and $L^{(5)}/2=6$, respectively, are halved due to the symmetry of Eq.~(\ref{eq:CorrelationFct}). Equations~(\ref{eq:RandomMoment2}) and (\ref{eq:RandomMoment4}) thus manifest the growth of measurement settings that is required for the evaluation of moments with increasing order, a fact that also yields interesting prospects for generalizations of spin-squeezing inequalities  derived in Refs.~\cite{SpinSqueez1,SpinSqueez2,SpinSqueez3,SpinSqueez4}. Lastly, note that in a similar manner one can obtain expressions for higher moments, but for the remainder of the paper we will mainly focus on $\mathcal R^{(2)}$ and $\mathcal R^{(4)}$.

\begin{figure}[t]
\begin{center}
\includegraphics[width=0.48\textwidth]{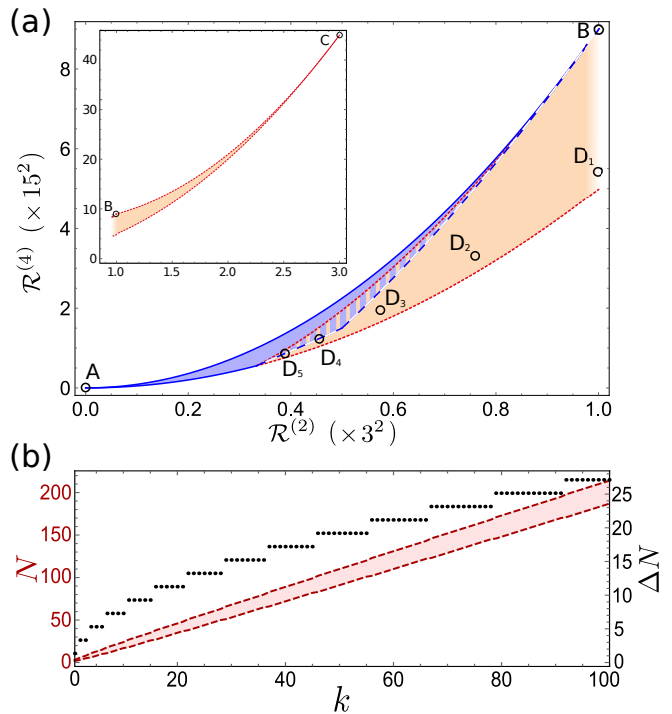}%
\end{center}
\caption{(a) Representation of the set of two-qubit separable (blue solid lines) and entangled (red dotted lines) BD states in the space spanned by the moments $\mathcal R^{(2)}$ and $\mathcal R^{(4)}$ in the range $0\leq \mathcal R^{(2)}\leq 1/3^2$. For $\mathcal R^{(2)}\leq1/3^3$ all states are separable, and for $1/3^3\leq \mathcal R^{(2)}\leq 1/3^2$ separable and entangled states have a non-zero overlap (striped region). The two-qubit criterion discussed in the main text is indicated by the white dashed curve. The inset depicts the rest of the set of entangled states in the range $1/3^2\leq \mathcal R^{(2)}\leq 1/3$. The maximally mixed state (A), the pure product states (B) and the Bell states (C) are indicated by labeled black circles. Circles labeled (D$_1$) to (D$_5$) represent Dicke reduced states $\ket{D^{N}_{k=2}}$, with $N=3,\ldots,7$. (b) Dicke states $\ket{D^{N}_{k}}$ detected from two-body correlations. The red area represents values of $N$ and $k$ for which the criterion (\ref{eq:2qubitcrit}) is violated. Black dots show the range $\Delta N=N_\text{max}-N_\text{min}$, where $N_\text{max}$ and $N_\text{min}$ indicate the piecewise parallel upper and lower bounds of the red area, as a function of $k$.  
} 
\label{fig_2}
\end{figure}

\textit{Bipartite entanglement.}--- An important subclass of two-qubit states is that of Bell diagonal (BD) states which are defined as 
$\rho_{\text{BD}}=\frac{1}{4}\big[\mathbb 1_4  +\sum_{j=x,y,z} c_j \sigma_j \otimes \sigma_j \big]$, with real parameters $c_j$, such that $0\leq |c_j|\leq 1$, and the eigenvalues $\lambda_j$ of $\rho_{\text{BD}}$ are given by $\lambda_{1,2}=(1\mp c_1 \mp c_2-c_3)/4$ and $\lambda_{3,4}=(1\pm c_1 \mp c_2+c_3)/4$ \cite{BellDiagStates}. In Fig.~\ref{fig_2} we present the set of BD states in the space spanned by the moments $\mathcal R^{(2)}$ and $\mathcal R^{(4)}$, obtained from an  analytic optimization over the parameters $c_1$, $c_2$ and $c_3$ (see App.~D).  In the same figure we indicate the division of the set of states into an separable and entangled part, as it results from the separability condition $\abs{c_1}+\abs{c_2}+\abs{c_3}\leq1$. Note that there remains a small overlap between the two sets containing both separable and entangled BD states which can be distinguished perfectly by taking into account also the moment $\mathcal R^{(6)}$, as shown in App.~D. Hence, the entanglement of Bell diagonal states is completely characterized by the first three non-vanishing moments $\mathcal R^{(t)}$, with $t=2,4,6$.

Further on, we note that for any general two-qubit state $\rho$ one can find a BD state $\rho_{\text{BD}}$ that has the same moments. This is a direct consequence of the fact that the $\mathcal R^{(t)}$'s are LU invariant and that the transformation eliminating the local Bloch vector components of $\rho$ is separable. 
In conclusion, the set of general and BD states are identical in the space spanned by the moments (see Fig.~\ref{fig_2}). Furthermore, as separable transformations are entanglement non-increasing, we obtain our main result: \textit{For separable states, one has}
\begin{align}
F(\mathcal R^{(2)},\mathcal R^{(4)})\geq 0,
\label{eq:2qubitcrit}
\end{align}
\textit{where $F$ is a piecewise polynomial function characterizing the border of the separable set, derived in App.~D}. It is evident that (\ref{eq:2qubitcrit}) detects more entangled states than any criteria depending only on either of the moments: $\mathcal R^{(2)}\leq 1/3^2$ or $\mathcal R^{(4)}\leq 1/5^2$ (see \cite{BriegelLUinv,tran1} and App.~C).

\textit{Multi-qubit entanglement.}--- As application of the above bipartite criterion for the detection of multi-qubit entanglement  we consider the  class of Dicke states which for $N$ qubits  read $\ket{D^{N}_k}=1/\sqrt{\binom{N}{k}}\sum_j P_j (\ket{1}^{\otimes k} \ket{0}^{\otimes (N-k)})$, where $k$ is the number of excitations and $\sum_j P_j$ denotes the sum over all non-equivalent permutations among the qubits. As Dicke states are invariant under permutations of their subsystems, we can detect their entanglement by focusing on any of their two-qubit marginals. We also emphasize that none of the states $\ket{D^{N}_k}$, for any $N$ and $k$, can be detected using only either of the moments $\mathcal R^{(2)}$ or $\mathcal R^{(4)}$. In contrast, our novel nonlinear criterion (\ref{eq:2qubitcrit}) is capable detecting Dicke state entanglement. For instance, in the case of the $N$-qubit $W$-state $\ket{W_N}=\ket{D^N_{k=1}}$ we can ascertain entanglement for  $N\leq 3$. The same holds for a subset of Dicke states with $k> 1$. In Fig.~\ref{fig_2}(b), we represent the set of Dicke states detected by our criterion for up to $200$ qubits. 

\begin{figure}[t]
\begin{center}
\includegraphics[width=0.49\textwidth]{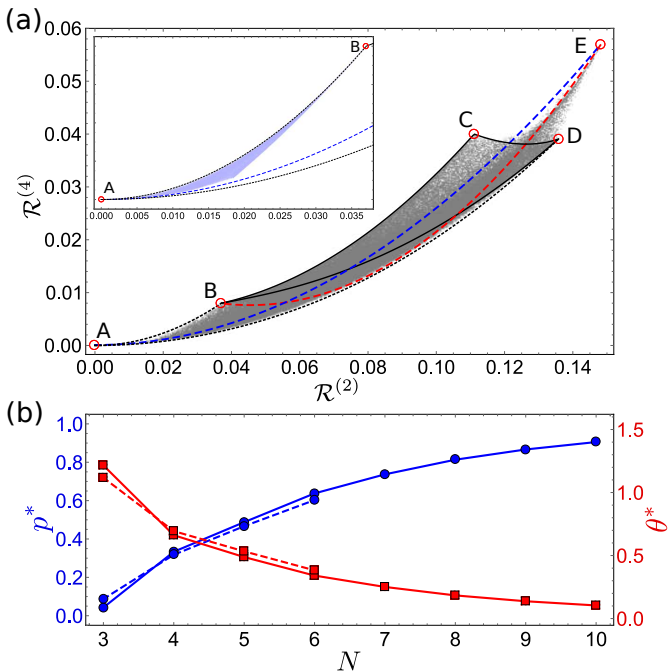}%
\end{center}
\caption{(a) Representation of the set of three-qubit states in the space spanned by the moments $\mathcal R^{(2)}$ and $\mathcal R^{(4)}$. Gray points correspond to randomly generated mixed quantum states. Labeled red circles indicate the maximally mixed state (A), all pure product states (B), bi-separable states of the form $\ket{\phi} \ket{\text{Bell}}$ (C), the three-qubit $W$- (D) and GHZ-state (E). While the  black lines connecting (B), (C) and (D) enclose the set $\mathcal W^{(3)}$, its mixed extension $\text{Conv}(\mathcal W^{(3)})$ is indicated by black dotted lines. The  noisy GHZ state (blue dashed line) and the pure state $\ket{\Psi(\theta)}$ (red dashed-dotted line) are shown for $0\leq p\leq 1$ and $0\leq\theta\leq \pi/2$, respectively. The inset shows randomly generated fully separable states (light blue points) in the range $0\leq \mathcal R^{(2)} \leq 1/3^3$. (b) Plot of the noise threshold $p^*$ (blue dots, left scale) up to which a GHZ state is detected to be not in $\text{Conv}(\mathcal W^{(N)})$. Solid and dashed lines refer to thresholds obtained from the criteria $\mathcal R^{(2)}\leq \chi^{(N)}$ and $m^{(N)} \mathcal R^{(2)}+\tilde b^{(N)}\geq \mathcal R^{(4)}$, respectively,  for varying number of qubits $N$. Accordingly, red squares show the amplitude threshold $\theta^*$ (red right scale) above which $\ket{\Psi(\theta)}$ is detected to be outside of $\mathcal W^{(N)}$.
}
\label{fig_3}
\end{figure}

In order to obtain better criteria that are capable of detecting more entangled states we have to take into account moments of $N$-body correlation functions. In this respect, we note that entanglement criteria based only on the second moment  have been subject of  investigations in the context of correlation tensor norms \cite{BriegelLUinv,JulioCorrMat1,JulioCorrMat2,JulioMarkus,ZukowskiRefFrame1,ZukowskiRefFrame2,LawsonRefFrame,KlocklHuber,tran1}. For instance, it is known that $\mathcal R^{(2)}\leq1/3^N$, for all separable $N$-qubit states. Here we ask whether these results can be improved upon  by combining $N$-body moments of different order. However, an analytical characterization of the borders of the set of (separable) states as presented for two qubits becomes very demanding already for three parties. Despite of this difficulty we gained insight  into the structure of the three-qubit state space by numerically generating more than $10^5$ random (fully separable) states. In Fig.~\ref{fig_3}(a) we present the results of this procedure. As for two qubits, we find that an advantage for entanglement detection is possible by taking into account $\mathcal R^{(4)}$. An analytical proof of this observation, in particular for more qubits, remains subject of future investigations.

\textit{Classes of  multipartite entanglement.}--- In the multipartite case it is also of interest to discriminate different classes of multi-partite entanglement which are defined through the concept of stochastic local operations and classical communication (SLOCC) \cite{SLOCC3qDuer,SLOCC3qAcin}. Two pure $N$-qubit states $\ket\Psi$ and $\ket\Phi$ are SLOCC equivalent if there exist invertible operations $A_i$, with $i=1,\ldots,N$, such that $\ket\Psi=\bigotimes_{i=1}^{N}A_i \ket\Phi$. The corresponding equivalence classes that result from this definition are referred to as SLOCC classes. While for three parties there exist two SLOCC classes of genuinely multipartite entangled states, the $W$- and the GHZ-class \cite{SLOCC3qDuer}, they become infinitely many already for $N=4$ \cite{Conny4qu}. In the following, we will concentrate our attention on the corresponding equivalence classes of $N$ qubits, referred to as $ \mathcal W^{(N)}$ and $ \mathcal{GHZ}^{(N)}$. As for separable states the respective  sets of mixed states are given by the convex hulls $\text{Conv}(\mathcal W^{(N)})$ and $\text{Conv}(\mathcal{GHZ}^{(N)})$ \cite{SLOCC3qAcin}.
 
For the characterization of the $W$-class it is helpful to resort to its standard form representing all pure $W$-class states up to LU transformations \cite{StForm3qAcin,StFormNq,StForm3qTurgut}. The latter allows us to numerically determine the borders of the pure $W$-class in the space spanned by the moments $\mathcal R^{(t)}$, as presented in Fig.~\ref{fig_3}(a), for $N=3$. In the same figure we present an estimate of the borders of the mixed $W$-class which has been obtained by minimizing over a subclass of $\text{Conv}(\mathcal W^{(N)})$ and confirmed by generating more than $10^5$ mixed $W$-class states (for details of this procedure see App.~E). As a result, we see that a discrimination of states outside of the $W$-class based on the knowledge of $\mathcal R^{(2)}$ and $\mathcal R^{(4)}$ is possible. 

With increasing qubit number the numerical characterization of the complete $W$-class becomes computationally more demanding. For this reason, we aim for simpler criteria which can be extended to larger numbers of qubits. 
One way of doing so is to compute the maximum of $\mathcal R^{(2)}$ in $\mathcal W^{(N)}$, and use its convexity to derive the criterion $\mathcal R^{(2)}\leq \max\limits_{\rho'\in \mathcal W^{(N)}} \mathcal R^{(2)}=:\chi^{(N)}$, for all $\rho\in \text{Conv}(\mathcal W^{(N)})$. 
We further test this criterion by applying it to the noisy GHZ state $p \mathbb 1+(1-p)\ketbra{\text{GHZ}}{\text{GHZ}}$ and a pure entangled state $\ket{\Psi(\theta)}=\cos{\theta} \ket{0}^{\otimes N}+\sin{\theta} \ket{1}^{\otimes N}$, and determine the corresponding noise and amplitude  thresholds, $p^*$ and $\theta^*$, respectively, up to which the states can be certified to be not in $\text{Conv}(\mathcal W^{(N)})$ (see Fig.~\ref{fig_3}(b)). Clearly, the performance of the criterion improves with growing qubit number. 

Lastly, we take an attempt to go beyond these results through a more general criterion that combines $\mathcal R^{(2)}$ and $\mathcal R^{(4)}$. Examining the structure of the $W$-class in Fig.~\ref{fig_3}(a) we expect that a line passing through the points C and D yields such an improvement. For arbitrary $N$, such a line can be derived by first maximizing individually, $\mathcal R^{(2)}$ and $\mathcal R^{(4)}$, over $\mathcal W^{(N)}$. The resulting arguments of this maximizations then allow us to define a line with slope $m^{(N)}$. Second, we maximize the $\mathcal R^{(4)}$-intercept $b^{(N)}$ of this line over $\mathcal W^{(N)}$ to ensure that it touches its border. Finally, as a linear combination of two convex functions with positive coefficients is again convex, we arrive at the criterion $m^{(N)} \mathcal R^{(2)}+\tilde b^{(N)}\geq \mathcal R^{(4)}$, for all $\rho\in \text{Conv}(\mathcal W^{(N)})$ and with $\tilde b^{(N)}=\max\limits_{\rho\in \mathcal W^{(N)}}  b^{(N)}$, as demonstrated in App.~F. The performance of the latter is presented Fig.~\ref{fig_3}(b). Evidently, the observed improvement for three qubits does not hold for larger qubit number. Hence, in order to improve the above results a more refined nonlinear witness is desirable.

\textit{Experimental considerations.}--- The discussed methods are of interest for photonic free-space quantum communication over distances of several hundreds of kilometers \cite{FreeSpaceQC}, which is currently in the process of being extended to space involving satellites orbiting the earth \cite{SatelliteQC1,SatelliteQC2,SatelliteQC3,SatelliteQC4}. Here, due to the motion, distance and number of involved satellites, the issue of sharing classical reference frames becomes particularly challenging \cite{StevePolytopes,LeandroRotInv,SteveAlignmentFreeComm}. 
In such a scenario, the moments (\ref{eq:RandomMoments}) can either be evaluated exactly through the fixed measurement settings involved in Eqs.~(\ref{eq:RandomMoment2}) and (\ref{eq:RandomMoment4}), in the spirit of RF independent entanglement citeria \cite{LawsonRefFrame}. Alternatively, they can be estimated using a statistical treatment based on a finite number of randomly chosen settings. The latter can be achieved experimentally by digitally generating random unitary transformations which are subsequently applied prior to the photon polarization measurements (see e.g. Ref.~\cite{RandUExp}). Also note that, according to the analysis presented in Refs.~\cite{tran1,tran2}, the number of random measurement settings needed to certify entanglement with confidence scales favourably with the system size, especially in the multipartite regime.

\textit{Conclusions.}--- A major challenge for the experimental detection and characterization of multipartite entanglement is the need of sharing a common RF allowing to coordinate measurements taken at a distance. In this work, we showed how to improve considerably existing techniques for RF independent entanglement verification by combining statistical moments of correlation functions obtained from measurements taken with randomly distributed settings. To this effect, we made use of designs which allow for a straightforward evaluation of the corresponding moments. 
We demonstrated the introduced techniques by applying them to detect entanglement in multi-qubit systems, and also to discriminate different classes of multipartite entanglement. 

Finally, our results yield interesting prospects for generalizations of the spin-squeezing inequalities derived in Refs.~\cite{SpinSqueez1,SpinSqueez2,SpinSqueez3,SpinSqueez4}. Although, such generalized inequalities would lose the LU invariance property, the extension of the number of  local measurement settings originating from the spherical $5$-design is expected to entail an improvement in their detection power.

\begin{acknowledgements}
We appreciate helpful comments by  H. Chau Nguyen on this work and we thank Nicolas Brunner, Marcus Huber, Lukas Knips, Daniel McNulty, Jasmin Meinecke, Tomasz Paterek, Gael Sent\'{\i}s, Timo Simnacher, Cornelia Spee and Xiao-Dong Yu for fruitful discussions. We acknowledge financial support from the ERC (Consolidator Grant 683107/TempoQ) and the DFG. NW acknowledges support from the House of Young Talents Siegen.
\end{acknowledgements}

\onecolumngrid

\section{A. Moments of random correlations}\label{app:A}
In the main text we focused on a system consisting of $N$ qubits. In a more general fashion one can consider $N$ $d$-dimensional systems (qudits) prepared in the initial state $\rho_\text{in}$ and subsequently measured locally in random bases $\{(\ket{u_n^{(0)}}= U_n\ket{0_n},\ket{u_n^{(1)}}= U_n\ket{1_n},\ldots,\ket{u_n^{(d-1)}}= U_n\ket{(d-1)_n}))\}_{n=1,\ldots,N}$, each specified by a random unitary transformation $U_n$ picked from the unitary group $\mathcal U(d)$. 
The generalization of the moments, defined in Eq.~(2) of the main text to qudits can then be defined as follows:
\begin{align}
\mathcal R^{(t)}&= \int_{\mathcal U(d)} d\eta(U_1) \ldots \int_{\mathcal U(d)} d\eta(U_N)
\expec{ (U_1\mathcal O U_1^\dagger\otimes \ldots \otimes U_N\mathcal O U_N^\dagger)}^t,
\label{app:RandomMomentsQudits}
\end{align}
where $t$ is a positive integer, $\eta$ the Haar measure on the unitary group $\mathcal U(d)$, and $\mathcal O$ describes an arbitrary observable diagonal in the computational basis $\{\ket{0_n},\ldots,\ket{(d-1)_n}\}$. Note that, while each $U_n\in \mathcal U(d)$ can be parametrized by $d^2$ angles, this number reduces in the case of the operators $U_n\mathcal O U_n^\dagger$ to $d(d-1)$ free parameters. Hence, we see that for $d=2$ the local measurement settings are characterized by two angles corresponding to the spherical coordinates fixing a position on the Bloch sphere $S^2$. Equation~(2) of the main text thus follows directly from Eq.~(\ref{app:RandomMoments}):
\begin{align}
\mathcal R^{(t)}&= \int_{\mathcal U(2)} d\eta(U_1) \ldots \int_{\mathcal U(2)} d\eta(U_N)
\expec{ (U_1\sigma_{z}U_1^\dagger\otimes \ldots \otimes U_N\sigma_{z}U_N^\dagger)}^t\\
&= \frac{1}{(4\pi)^N} \int_{S^{2}} d\boldsymbol u_1\ldots \int_{S^{2}} d\boldsymbol u_N  E(\boldsymbol u_1,\ldots,\boldsymbol u_N)^t.
\label{app:RandomMoments}
\end{align}
In particular, for $d=2$ it is easy to see that all odd moments are zero due to the symmetry of the correlation functions $E(\boldsymbol u_1,\ldots,\boldsymbol u_N)$ with respect to a reflection on the Bloch sphere: $E(\boldsymbol u_1,\ldots,-\boldsymbol u_i,\ldots, \boldsymbol u_N)=-E(\boldsymbol u_1,\ldots,\boldsymbol u_N)$. 

In the following, we show that the moments $\mathcal R^{(t)}$ can be calculated using unitary $t$-designs (or spherical $t$-designs for $d=2$), rather than averaging over the whole unitary group. 

\section{\label{app:B} B. Designs}
\subsection{\label{app:B1} B.1. Unitary designs}
Let us denote by $\mathrm{Hom}(r,s)$ the set of all homogeneous polynomials $P_{r,s}(U)$, with support on the space of unitary matrices $\mathcal U(d)$, that is of degree at most $r$ and $s$, respectively, in each of the matrix elements of $U$ and their complex conjugates. For example, the polynomial $P_{r=2,s=2}(U,V)=U^\dagger V^\dagger U V$ is of degree $r=s=2$ in the matrices $U$ and its conjugate ones. 
With this we arrive at the following definition:
\begin{definition}[\text{Unitary t-designs}]
A {unitary $t$-design} is a set of unitary matrices $\{U_k|k=1,\ldots,K^{(t)}\}\subset \mathcal U(d)$, with cardinality $K^{(t)}$, such that 
\begin{align}
\frac{1}{K^{(t)}} \sum_{k=1}^{K^{(t)}} P_{t,t}(U_k) =\int_{\mathcal U(d)} P_{t,t}(U) d\eta(U),
\end{align} 
for all homogeneous polynomials $P_{t,t}\in \mathrm{Hom}(t,t)$ of degree smaller or equal than $t$, and where $\eta(U)$ denotes the normalized Haar measure.
\end{definition}
Equivalent definitions of unitary $t$-designs which will be used later on are given through the following theorem \cite{Dankert}:
\begin{theorem}[Twirling]\label{theorem1}
Given a unitary $t$-design $\{U_k|k=1,\ldots,K^{(t)}\}\subset \mathcal U(d)$, the following two statements hold:
\begin{itemize}
\item[(1)] For all $A\in \mathcal B(\mathcal H^{\otimes t})$:
\begin{align}
\frac{1}{K^{(t)}} \sum_{k=1}^{K^{(t)}} U_k^{\otimes t}A (U_k^{\otimes t})^\dagger =\int_{\mathcal U(d)} U^{\otimes t}A (U^{\otimes t})^\dagger d\eta(U).
\end{align} 

\item[(2)] For all channels $\Lambda_1,\ldots,\Lambda_{n-1} : \mathcal B(\mathcal H)\rightarrow \mathcal B(\mathcal H)$:
\begin{align}
\frac{1}{K^{(t)}} \sum_{k=1}^{K^{(t)}} T^{(t-1)}_{ \Lambda_{U_k},\{\Lambda_k\}_{k=1}^{n-1}}(A)=\int_{\mathcal U(d)}d\eta \  T^{(t-1)}_{ \Lambda_U,\{\Lambda_k\}_{k=1}^{n-1}}(A),
\label{eq:TwirlingChannel}
\end{align} 
where the $T^{(t-1)}_{\Lambda_U,\{\Lambda_k\}_{k=1}^{n-1}}$ denotes a $(t-1)$-fold twirling of the channels $\Lambda_1,\ldots,\Lambda_{n-1}$ with the unitary transformation $\Lambda_U(\cdot)= U \cdot  U^\dagger$, defined as:
\begin{align}
T^{(t-1)}_{\Lambda_U,\{\Lambda_k\}_{k=1}^{n-1}}(A):= \Lambda_U^\dagger \circ \Lambda_{t-1} \circ \Lambda_U \circ \Lambda_{t-2} \circ \Lambda_U^\dagger\circ \ldots \circ\Lambda_U \circ   \Lambda_2 \circ \Lambda_U^\dagger \circ   \Lambda_1 \circ \Lambda_U(A),
\label{eq:Twirling}
\end{align}
where $\Lambda_U^\dagger(\cdot)= U^\dagger \cdot  U$ denotes the adjoint transformation of $\Lambda_U$, and $A\in\mathcal B(\mathcal H)$.
\end{itemize}
\end{theorem}

An example of a unitary $3$-design is given by the Clifford group \cite{Cliff3Design,CliffNo4Design}. In the case of a single qubit the Clifford group has $24$ elements which can be generated from the Hadamard gate $ H$ and the phase gate $ S=e^{i \frac{\pi}{4} \sigma_z}$. Further on, in Ref.~\cite{Gross5Design} the existence of a qubit $5$-design of one qubit was noted. The latter is given by the unitary representation of the special linear group $SL(2,\mathbb F_5)$ of invertible $2\times 2$ matrices over the finite field $\mathbb F_5$ with five elements. In the following, we shortly outline how to generate this design.

First, we extract a set of generators of $SL(2,\mathbb F_5)$ from the GAP character library using the package REPSN, as outlined in \cite{Gross5Design}, leading to:
\begin{align}
\begin{pmatrix}
-1 & 0 \\
0 & -1 
\end{pmatrix},
\begin{pmatrix}
\omega^{10} &\omega^{11} + \omega^{14}\\
-\omega^2 - 
    \omega^8 & -\omega^{10} 
\end{pmatrix},
\begin{pmatrix}
-\omega^{11} - \omega^{14} & 
  \omega^6 + \omega^9 \\
  -\omega - \omega^2 - \omega^4 - 
   \omega^7 - \omega^8 - \omega^{13} & 
  \omega^{11} + \omega^{14}
\end{pmatrix},
\begin{pmatrix}
0& \omega^5\\
-\omega^{10} & -\omega^3 - \omega^{17} 
\end{pmatrix},
\label{eq:Generators5Design}
\end{align}
with $\omega=e^{i 2\pi/15}$. At this point we note that the generators given in Table~III of Ref.~\cite{Gross5Design} contain a typo. Next, with the matrices (\ref{eq:Generators5Design}) we generate the $120$ group elements of $SL(2,\mathbb F_5)=\{S_k|k\in\{1,\ldots,120\}\}$ from which we can then extract an appropriate unitary representation using the transformation:
\begin{align}
U_k=\sqrt{P} S_k \sqrt{P}^{-1},
\end{align}
where 
\begin{align}
P=\sum_{k=1}^{120} S_k^\dagger S_k>0.
\end{align}
Further on, after eliminating all matrices that are equal up to a global phase $e^{i\phi}$, we end up with a set of $60$ unitary matrices representing the corresponding unitary $5$-design.

\subsection{\label{app:B} B.2. Spherical designs}
In Section A we saw that in the case of systems consisting of qubits the evaluation of the moments $\mathcal R^{(t)}$ boils down to an integration over the local Bloch spheres $S^2$. In this case, instead of using a unitary $t$-design to evaluate the respective moments, we can also resort to the concept of spherical designs. In general, a spherical $t$-design consist of a finite set of points $\{\boldsymbol u_k\in  S^{2}|k=1,\ldots,L^{(t)}\}\subset S^{2}$  fulfilling the property 
\begin{align}
\frac{1}{L^{(t)}} \sum_{k=1}^{L^{(t)}} P_{t}(\boldsymbol u_k) =\int_{S^{2}}d\boldsymbol u \ P_{t}(\boldsymbol u),
\label{eq:tDesignDef}
\end{align} 
for all homogeneous polynomials $P_{t}:S^2\rightarrow \mathbb R$ of degree at most $t$. It thus suffices to resort to spherical $t$-designs as long as one is interested in calculating averages of polynomials of degree at most $t$ over the Bloch sphere $S^2$. 

\begin{figure}[t]
\begin{center}
\includegraphics[width=0.4\textwidth]{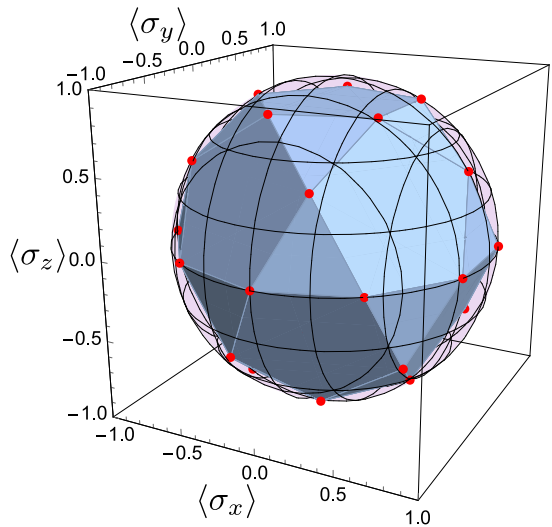}%
\end{center}
\caption{Plot of the $15$ measurement settings corresponding to the spherical $5$-design generated from the unitary $5$-design $SL(2,\mathbb F_5)$. Each direction yields two points on the Bloch sphere which together yield the vertices of an icosidodekahedron, i.e. a polyhedron with twenty triangular faces and twelve pentagonal faces.}
\label{app:fig_1}
\end{figure}

One way to generate spherical designs is to extract them from unitrary designs. For instance, by applying the elements of the single qubit Clifford group to one of the Pauli matrices, e.g. $\sigma_z$, we are left with the following set of inequivalent operators $\{\pm \sigma_x,\pm\sigma_y,\pm\sigma_z\}$. The latter correspond to the following set of unit vectors $\{\pm\boldsymbol e_i|i=x,y,z\}$ which form a spherical $3$-design. Similarly, one can generate a spherical $5$-design from the $60$ element unitary $5$-design $SL(2,\mathbb F_5)$. To do so, we calculate again all inequivalent directions on the Bloch sphere originating from the operators $\sigma_{\boldsymbol u_k}=U_k\sigma_zU_k^\dagger$, for all elements $U_k$ of the unitary $5$-design. The result is a set of 30 vertices on the Bloch sphere forming an icosidodekahedron (see Fig.~\ref{app:fig_1}). 

\begin{figure}[t]
\begin{center}
\includegraphics[width=0.44\textwidth]{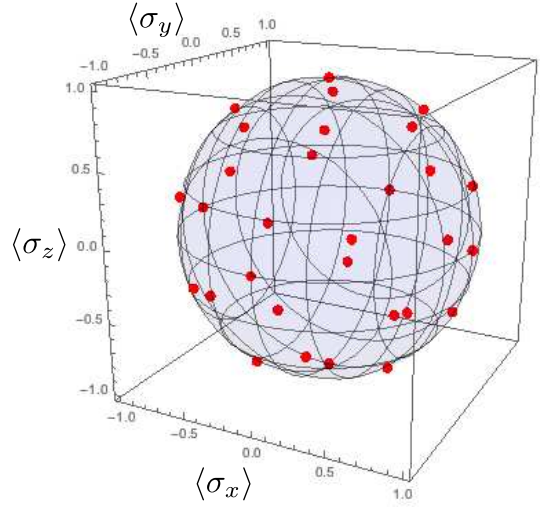}%
\end{center}
\caption{Plot of the $30$ measurement settings corresponding to the spherical $7$-design found in \cite{ExamplesSphericalDesigns}.}
\label{app:fig_2}
\end{figure}

The advantage of spherical designs is that they are easier to find, as compared to unitary designs. Indeed, instead of constructing spherical designs from unitary ones, one can search them also directly by checking the relation~(\ref{eq:tDesignDef}) for sets of vertices on the sphere $S^2$. Such a search was carried out in Ref.~\cite{ExamplesSphericalDesigns}. In particular, they found spherical designs on the $2$-sphere $S^2$ for $t$'s up to 20 consisting of up to 100 elements. For instance, they showed that the regular icosahedron (see Fig.~1(b) of the main text), consisting of 12 vertices, already forms a $5$-design. Furthermore, one can find a spherical $7$-design consisting of 30 vertices on $S^2$ (see Fig.~\ref{app:fig_2}). Interestingly, the icosidodekahedron, presented in Fig.~\ref{app:fig_1}, has the same number of vertices but does not even constitute a $6$-design.  Hence, the relation between spherical and unitary designs is not straightforward.

Lastly, we emphasize that for systems of larger local dimensions it is not possible to resort to spherical designs instead of unitary ones for the evaluation of the moments $\mathcal R^{(t)}$. This is due to the fact that the space parametrizing all possible measurement settings of a qudit, i.e. all possible observables of the form $U\mathcal O U^\dagger$ (see Sec. A), is not in one to one correspondence with points on a generalized Bloch sphere $S^{d^2-1}$. It rather forms a submanifold of $S^{d^2-1}$ characterized by $d(d-1)$ parameters. Hence, generalized spherical designs on $S^{d^2-1}$ will not be usefull in this case. 

\section{\label{app:C}C: Random moments from designs}
In this section we show that the $t$-th power of the correlation function $\expec{ (U_1\mathcal O U_1^\dagger\otimes \ldots \otimes U_N\mathcal O U_N^\dagger)}$ is a polynomial of degree $t$ in the entries of the local random unitaries $U_n\in \mathcal U(d)$, and their complex conjugates, and thus that the moments $\mathcal R^{(t)}$ can be evaluated over a unitary $t$-design instead of the whole unitary group $\mathcal U(d)$. To do so, we first introduce a map $\Phi: \mathcal B((\mathbb C^d)^{\otimes N})\rightarrow \mathcal B((\mathbb C^d)^{\otimes N})$ which is defined through
\begin{align}
\Phi (A)=\mathcal O\otimes\ldots \otimes \mathcal O \mathrm{tr}[\rho A],
\end{align}
and where $\rho\in \mathcal B((\mathbb C^d)^{\otimes N})$ is an arbitrary quantum state. Next, we consider the $(t-1)$-fold twirling of the map $\Phi$ with the local unitary transformation $\Lambda_{U_1,\ldots,U_N}(\cdot)=(U_1\otimes \ldots \otimes U_N )(\cdot) (U_1\otimes \ldots \otimes U_N )^\dagger$, which yields
\begin{align}
T_{\Lambda_{U_1,\ldots,U_N}, \Phi }^{(t-1)}(\mathcal O^{\otimes N}) =\Lambda_{U_1,\ldots,U_N}(\mathcal O^{\otimes N}) \text{tr}[\rho \Lambda_{U_1,\ldots,U_N}(\mathcal O)]^{t-1},
\end{align}
and take its expectation value to arrive at
\begin{align}
\text{tr} [\rho T_{\Lambda_{U_1,\ldots,U_N}, \Phi }^{(t-1)}(\mathcal O^{\otimes N}) ]=\expec{ (U_1\mathcal O U_1^\dagger\otimes \ldots \otimes U_N\mathcal O U_N^\dagger)}^t.
\end{align}
At this point it is a straightforward consequence of Theorem~\ref{theorem1} that
\begin{align}
\int_{\mathcal U(2)}\ldots \int_{\mathcal U(2)} \expec{T_{\Lambda^{(N)}_{U_1,\ldots,U_N}, \Phi }^{(t-1)}(\rho)} d\eta(U_1)\ldots d\eta(U_N)=\frac{1}{(K^{(t)})^N}\sum_{k_1,\ldots,k_N=1}^{K^{(t)}}  \expec{T_{\Lambda_{U_{k_1},\ldots,U_{k_N}}, \Phi }^{(t-1)}(\rho)},
\label{eq:TwirlingRandCorr}
\end{align}  
and thus
\begin{align}
\mathcal R^{(t)}&=\frac{1}{(K^{(t)})^N}\sum_{k_1,\ldots,k_N=1}^{K^{(t)}}\expec{ (U_{k_1}\mathcal O U_{k_1}^\dagger\otimes \ldots \otimes U_{k_N}\mathcal O U_{k_N}^\dagger)}^t.
\label{eq:TwirlingRandCorrDesign}
\end{align}  
We thus showed that the moments (\ref{app:RandomMoments}) can be evaluated using unitary $t$-designs instead of carrying out the integrals over $\mathcal U(d)$. 

In the case of qubits ($d=2$) Eq.~(\ref{eq:TwirlingRandCorrDesign}) reduces to Eq. (4) of the main text:
\begin{align}
\mathcal R^{(t)}&=\frac{1}{(K^{(t)})^N}\sum_{k_1,\ldots,k_N=1}^{K^{(t)}}\expec{U_{k_1}^\dagger\sigma_z U_{k_1}\otimes\ldots \otimes U_{k_N}^\dagger\sigma_z U_{k_N}}^t \nonumber \\
&=\frac{1}{(L^{(t)})^N} \sum_{k_1,\ldots,k_N=1}^{L^{(t)}} \expec{\sigma_{\boldsymbol u_{k_1}}\otimes \ldots \otimes \sigma_{\boldsymbol u_{k_N}}}^t,
\label{eq:TwirlingRandCorrDesign}
\end{align}  
where $L^{(t)}\leq K^{(t)}$ denotes the count of the remaining non-equivalent measurement directions $\{\boldsymbol u_i\}_{i=1,\ldots,L^{(t)}}$, after dropping those for which it exist $j$'s such that $\sigma_{\boldsymbol u_{k_i}}=\sigma_{\boldsymbol u_{k_j}}$.
Hence, in order to calculate the moments $\mathcal R^{(t)}$ for systems of qubits it is enough to average  $\left[E(\boldsymbol u_1,\ldots,\boldsymbol u_N)\right]^t$ over a finite number $L^{(t)}$ of nonequivalent Bloch sphere directions $\{\boldsymbol u_{i}|i=1,\ldots,L^{(t)}\}\subset S^2$. The set of vertices $\{\boldsymbol u_{i}|i=1,\ldots,L^{(t)}\}\subset S^2$ is then called a spherical $t$-design because the function $\left[E(\boldsymbol u_1,\ldots,\boldsymbol u_N)\right]^t$ is a polynomial of degree $t$ on $S^2$. The last point is easy to see because the correlation function $E(\boldsymbol u_1,\ldots,\boldsymbol u_N)=\sum_{j_1,\ldots,j_N=x,y,z} u_1^{(j_1)}\cdot \ldots\cdot u_N^{(j_N)} \expec{\sigma_{j_1}\otimes \ldots \otimes \sigma_{j_N}}$, is a linear function of the components of the $\boldsymbol u_n$'s. Hence, using the $6$ element spherical $3$-design and $12$ element spherical $5$-design, presented in Sec.~B.2., we arrive at the Eqs.~(5) and (6) of the main text, respectively. Note that in Eqs.~(5) and (6) the number of summands is $L^{(t)}/2$ because for even $t$ we can drop the respective anti-parallel settings. Lastly, we remark that one could in principle use the spherical $5$-design to calculate both moments, as every $t$-design is also a $(t-1)$-design, but it is desirable to minimize the number of settings as much as possible as possible.

As an example we consider the case of a single qubit. For $N=1$ Eqs.~(5) and (6) become:
\begin{align}
\mathcal R_{N=1}^{(2)}&=\frac{1}{3} \sum_{i=x,y,z} \expec{\sigma_{i}}^2\leq \frac{1}{3} ,\label{eq:RandomMoment2qu1}  \\
\mathcal R_{N=1}^{(4)}&=\frac{1}{6} \sum_{i=1}^{6} \expec{\sigma_{\boldsymbol v_{i}}}^4\leq \frac{1}{5}.
\label{eq:RandomMoment4qu1}
\end{align}
The upper bounds can be straightforwardly calculated using the fact that $\mathcal R_{N=1}^{(2)}$ and $\mathcal R_{N=1}^{(4)}$ are invariant under local unitary transformations and convex in the state $\rho$. Equations~(\ref{eq:RandomMoment2qu1}) and (\ref{eq:RandomMoment4qu1}) yield entanglement criteria $\mathcal R^{(2)}\leq 1/3^N$ and $\mathcal R^{(4)}\leq 1/5^N$, respectively, using a standard proof as presented, for instance, in \cite{KlocklHuber}. While the former criterion is known \cite{BriegelLUinv,KlocklHuber} the latter is novel. Nevertheless, we did not find an advantage in using one over the other. To obtain an advantage for entanglement detection the  moments $\mathcal R^{(2)}$ and $\mathcal R^{(4)}$ have to be combined, as discussed in the following.

\section{ D: Two-qubit entanglement detection}\label{app:D}

Bell diagonal states read  $\rho_{\text{BD}}=\frac{1}{4}\big[\mathbb 1_4  +\sum_{j=1}^3 c_j \sigma_j \otimes \sigma_j \big]$, with real parameters $c_j$, such that $0\leq |c_j|\leq 1$, and the corresponding eigenvalues $\lambda_{1,2}=(1\mp c_1 \mp c_2-c_3)/4$ and $\lambda_{3,4}=(1\pm c_1 \mp c_2+c_3)/4$. Bell diagonal states are separable iff $|c_1|+|c_2|+|c_3|\leq 1$. In the following we will show that this separability criteria can be directly transferred to the first three nonvanishing moments $\mathcal R^{(t)}$, with $t=2,4,6$. We start, however, by focusing on the moments $\mathcal R^{(2)}$ and $\mathcal R^{(4)}$ only as they have been at the core of the discussion in the main-text. 

\subsection{D.1. Detection of Bell-Diagonal states based on $\mathcal R^{(2)}$ and $\mathcal R^{(4)}$}\label{app:D.1}

Direct evaluation of Eq.~(5) and (6) of the main text for Bell diagonal states yields:
\begin{align}
\mathcal R^{(2)}&=\frac{1}{9} (c_1^2 + c_2^2 + c_3^2),\label{eq:Moment2BDS}\\
\mathcal R^{(4)}&=\frac{2}{75} (c_1^4 + c_2^4 + c_3^4)+\frac{27}{25} (\mathcal R^{(2)})^2,\label{eq:Moment4BDS}
\end{align}
which are polynomials of degree smaller equal than four in the coefficients $c_i$, with $i=1,2,3$. In order to determine the borders of the set of (separable) states our aim is to maximize (minimize) one of the moments with respect to the other while keeping the condition $|c_1|+|c_2|+|c_3|=x$ fixed.  Then, after the transformation $y:=3^2 \mathcal R^{(2)}$ and $z:=15^2 \mathcal R^{(4)}/6-y^2/2$, we arrive at the following system of equations:
\begin{align}
x&=|c_1|+|c_2|+|c_3|,\label{eq:SystEqsX}\\
y&=c_1^2+c_2^2+c_3^2,\label{eq:SystEqsY}\\
z&=c_1^4+c_2^4+c_3^4,\label{eq:SystEqsZ}
\end{align}
which can be solved algebraically. Note that when doing so we can restrict ourselves to the condition $c_1+c_2+c_3=x$, with $c_i\geq0$, as long as $y\leq x/2$, due to the symmetry of the (separable) set of Bell diagonal states in the parameter space $(c_1,c_2,c_3)$. As long as this holds the intersection of the ball defined through Eq.~(\ref{eq:SystEqsY}) and the octahedron defined through Eq.~(\ref{eq:SystEqsX}), in the space spanned by the $c_i$'s, is smaller than each of the faces of the octahedron, i.e. it is given by a circle. For $x/2<y<x$ we can also stick to the condition $c_1+c_2+c_3=x$ but we have to add additional constraints limiting the borders of the corresponding face of the octahedron: $c_1+c_3\leq x$, $ c_1+c_2\leq x$, $c_2+c_3\leq x$. 

In general, after having solved the above system of equations we simply have to check for which of the given solutions $R^{(4)}=(6 z+ 3 y^2)/15^2$ is maximal or minimal and that the chosen solution fulfils the positivity constraints of the corresponding eigenvalues of the Bell-diagonal states. This procedure then yields the corresponding parametrization of the set of (separable) states  by imposing the constraints $x\leq3$ ($x\leq 1$). Hence, all in all we find that the border of the set of states (represented in Fig.~2(a) of the main text) is lower bounded by
\begin{align}
f_{\text{LB}}(\mathcal R^{(2)})=\frac{5\times 3^4}{15^2} (\mathcal R^{(2)})^2,
\end{align}
and upper bounded by 
\begin{align}
f_{\text{UB}}(\mathcal R^{(2)})=\left\{
                \begin{array}{lll}
                   \frac{9\times 3^4}{15^2} (\mathcal R^{(2)})^2,&\text{for} &0\leq \mathcal R^{(2)}\leq 1/3^2 , \\ \\
                  \frac{3^2}{15^2}
                   (1 - 6 \mathcal R^{(2)} + 2\times 3^3 (\mathcal R^{(2)})^2), &\text{for} &\frac{1}{3^2}\leq \mathcal R^{(2)}\leq 1/3.
                \end{array}
                \right.
\end{align}
In the same way we find that the set of separable states (see Fig.~2(a) of the main text in blue) is lower bounded by:
\begin{align}
f_{\text{LB},\text{sep}}(\mathcal R^{(2)})=\left\{
                \begin{array}{lll}
                   \frac{5\times 3^4}{15^2} (\mathcal R^{(2)})^2,&\text{for} &0\leq \mathcal R^{(2)}\leq 1/3^3 , \\ \\
                  \frac{3^2}{15^2}\left(2 \mathcal R^{(2)}+{6\ 3^2 (\mathcal R^{(2)})^2}-\frac{4 \sqrt{2}}{3^4} \sqrt{27 \mathcal R^{(2)}-1}^{3}-\frac{7}{3^4}\right), &\text{for} &1/3^3\leq \mathcal R^{(2)}\leq 1/(2\times3^2),\\ \\
                  \frac{3^2}{15^2}\left(54 (\mathcal R^{(2)})^2+6 \mathcal R^{(2)}-\frac{1}{3}\right), &\text{for} &1/(2\times3^2)\leq \mathcal R^{(2)}\leq 1/3.
                \end{array}
                \right.
\end{align}
and upper bounded by $f_{\text{UB},\text{sep}}(\mathcal R^{(2)})=f_{\text{UB}}(\mathcal R^{(2)})$, for $0\leq \mathcal R^{(2)}\leq 1/3^2$. Finally, the set of entangled states (represented in Fig.~2(a) in red) is lower bounded by $f_{\text{LB},\text{ent}}(\mathcal R^{(2)})=f_{\text{LB}}(\mathcal R^{(2)})$, for $0\leq \mathcal R^{(2)}\leq 1/3^2$, and upper bounded by 
\begin{align}
f_{\text{UB},\text{ent}}(\mathcal R^{(2)})=\left\{
                \begin{array}{lll}
                   \frac{9\times 3^4}{15^2} (\mathcal R^{(2)})^2,&\text{for} &0\leq \mathcal R^{(2)}\leq 1/3^2 , \\ \\
                  \frac{3^2}{15^2}\left(2 \mathcal R^{(2)}+{6\ 3^2 (\mathcal R^{(2)})^2}+\frac{4 \sqrt{2}}{3^4} \sqrt{27 \mathcal R^{(2)}-1}^{3}-\frac{7}{3^4}\right), &\text{for} &1/3^2\leq \mathcal R^{(2)}\leq 1/3.
                \end{array}
                \right.
\end{align}
In conclusion, we find that the function $F(\mathcal R^{(2)},\mathcal R^{(4)})$ defining the two-qubit criterion $F(\mathcal R^{(2)},\mathcal R^{(4)})\geq 0 $ in the main text reads: $F(\mathcal R^{(2)},\mathcal R^{(4)})=\mathcal R^{(4)}-f_{\text{LB},\text{sep}}(\mathcal R^{(2)})$.

\subsection{\label{app:D.2}D.2. Detection of Bell-Diagonal states based on $\mathcal R^{(2)}$, $\mathcal R^{(4)}$ and $\mathcal R^{(6)}$}

Here we show that the overlap of the area representing separable and entangled states in Fig.~2(a) is lifted in the three-dimensional space spanned by $\mathcal R^{(2)}$, $\mathcal R^{(4)}$ and $\mathcal R^{(6)}$. To do so, we first have to evaluate the sixth moment for Bell diagonal states which can be straightforwardly  done  using the $30$ element spherical $7$-design presented in Sec.~B.2. Here, however, we do so by directly evaluating the integrals in Eq.~(\ref{app:RandomMoments}) which is feasible due to the small number of parameters involved in the Bell-diagonal states. 

Plugging the definition of Bell diagonal states into the formula of the $t$-th random moment (\ref{app:RandomMoments}), for $N=2$, thus yields:
\begin{align}
\mathcal R^{(t)}=\frac{1}{(4\pi)}\sum_{t=n_1+n_2+n_3} \frac{t!}{n_1!n_2!n_3!} c_1^{n_1}c_2^{n_2}c_3^{n_3} \left(\int_0^\theta d\theta (\sin{\theta})^{n_1+n_2+1} (\cos{\theta})^{n_3}\right)^2 \left( \int_0^{2\pi} d\phi (\cos{\phi})^{n_1} (\sin{\phi})^{n_2} \right)^2,
\label{eq:MomentsBellDiagSts}
\end{align}
where we have used the multinomial formula $(x_1 + \cdots + x_m)^n
 = \sum\limits_{k_1+k_2+\cdots +k_m = n} \frac{n!}{k_1!\cdot \ldots \cdot k_m!}
 x_1^{k_1} x_2^{k_2} \cdots x_m^{k_m}$. The integrals contained in Eq.~(\ref{eq:MomentsBellDiagSts}) can be directly solved:
\begin{align}
 \int_0^{\pi} d\theta (\cos{\phi})^{n_1} (\sin{\phi})^{n_2} &= \frac{(1+(-1)^{n_1})\Gamma(\frac{1+n_1}{2})\Gamma(\frac{1+n_2}{2})}{2\Gamma(\frac{1}{2}(2+n_1+n_2))},\\
\int_0^{2\pi} d\phi (\cos{\phi})^{n_1} (\sin{\phi})^{n_2} &= \frac{(1+(-1)^{n_1})(1+(-1)^{n_1+n_2}) \Gamma(\frac{1+n_1}{2})\Gamma(\frac{1+n_2}{2})}{2\Gamma(\frac{1}{2}(2+n_1+n_2))},
\end{align} 
for $n_1,n_2>0$ and where $\Gamma(x)$ denotes the Euler gamma function. The sixth moment thus becomes: 
\begin{align}
\mathcal R^{(6)}&=\frac{8}{735} (c_1^6 + c_2^6 + c_3^6)-\frac{486}{245} (\mathcal R^{(2)})^3+\frac{135}{49} \mathcal R^{(2)} \mathcal R^{(4)}.
\label{eq:6thMomentsBellDiag}
\end{align}
Given Eq.~(\ref{eq:6thMomentsBellDiag}), we can simply plug in the solution of Eqs.~(\ref{eq:SystEqsX})-(\ref{eq:SystEqsZ}) and verify that in the space spanned by $\mathcal R^{(2)}$, $\mathcal R^{(4)}$ and $\mathcal R^{(6)}$, the set of separable and entangled states do not overlap anymore. In particular, we find that the separating border is parametrized through the function:
\begin{align}
g(\mathcal R^{(2)},\mathcal R^{(4)})=&\frac{1}{1960}\left(26244 (\mathcal R^{(2)})^4-17496 (\mathcal R^{(2)})^3-24300 (\mathcal R^{(2)})^2 \mathcal R^{(4)}\right.\nonumber \\
&\left.+13500 \mathcal R^{(2)} \mathcal R^{(4)}-36 \mathcal R^{(2)}+5625 (\mathcal R^{(4)})^2+150 \mathcal R^{(4)}+1\right).
\end{align}
Hence, if we find a BD state with $\mathcal R^{(6)}\leq  g(\mathcal R^{(2)},\mathcal R^{(4)})$ ( $\mathcal R^{(6)}> g(\mathcal R^{(2)},\mathcal R^{(4)})$ ) we can conclude that the state is separable (entangled).

\subsection{\label{app:D.3} D.3: General two-qubit states}%

For completeness, we show here that the introduced criterion $F(\mathcal R^{(2)}_{\rho_{sep}},\mathcal R^{(4)}_{\rho_{sep}})\geq 0 $, for Bell-diagonal states, also yields a sufficient entanglement criterion for general two-qubit states. Hence, we have  to show that $F(\mathcal R^{(2)}_{\rho_{sep}},\mathcal R^{(4)}_{\rho_{sep}})\geq 0 $, for all separable two-qubit states $\rho_{sep}$. For clarity we indicate the state for which the $\mathcal R^{(t)}$'s are evaluated as a subscript. 

Let us start with a general two-qubit density matrix:
\begin{align}
\rho=\frac{1}{4}\big[\mathbb 1_4  + (\boldsymbol v^{(A)}\cdot \boldsymbol\sigma )\otimes \mathbb1_2+\mathbb 1_2 \otimes (\boldsymbol w^{(A)}\cdot \boldsymbol\sigma)+\sum_{i,j=1}^3 c_{i,j} \sigma_i \otimes \sigma_j \big],
\end{align}
where $\boldsymbol v^{(A)}$ and $\boldsymbol w^{(B)}$ are the Bloch vectors of the reduced subsystems A and B, $C=\{c_{i,j}\}_{i,j=1,\ldots,3}$ denotes the correlation matrix, and $\boldsymbol\sigma=(\sigma_x,\sigma_y,\sigma_z)^{\top}$. Note that the application of a local unitary transformation $U_A\otimes U_B$ to the state $\rho$ leads to a transformation $O_ACO_B^\top$ of the corresponding correlation matrix $ C$ with orthogonal matrices $O_A$ and $O_B$. We can thus always find a local unitary transformation such that $O_A'CO_B'^\top=\text{diag}(c_{1},c_2,c_3)$, and $\tilde{\boldsymbol v}^{(A/B)}=O_{A/B}'\boldsymbol v^{(A/B)}$. Further on, we apply the separable transformation $\rho\rightarrow \frac{1}{4}(\rho+\sum_{i=x,y,z}\sigma_i\otimes\sigma_i\rho\sigma_i\otimes\sigma_i)$, which eliminates the local Bloch vector components $\tilde{\boldsymbol v}^{(A)}$ and $\tilde{\boldsymbol w}^{(B)}$, and leaves us with the state $\tilde\rho=\frac{1}{4}\left(\mathbb 1_4+\sum_{i=1}^3 c_i\sigma_i\otimes\sigma_i\right)$. Lastly, as the $\mathcal R^{(t)}$'s depend only on two-body correlation functions and are LU invariant we have $\mathcal R_{\rho}^{(t)}=\mathcal R_{\tilde\rho}^{(t)}$, for all $t\in \mathbb N$. Hence, we have transformed $\rho$ into $\tilde\rho$ with an entanglement non-increasing transformation that preserves the moments $\mathcal R^{(t)}$, and thus shows that $F(\mathcal R^{(2)}_{\rho_{sep}},\mathcal R^{(4)}_{\rho_{sep}})\geq 0 $ is a sufficient entanglement criterion.

\section{\label{app:E} E: Multi-qubit entanglement detection}
\subsection{\label{app:E.1}E.1: With two-body moments}
In this section we check the capability of the bipartite entanglement criterion derived in Sec.~D.1 for multi-qubit entanglement detection. We focus first on entanglement criteria that are based on moments of two-body correlation functions only. We thus define:
\begin{align}
\mathcal R_{\alpha,\beta}^{(t)}&= \frac{1}{(4\pi)^2} \int d\boldsymbol u_\alpha \int d\boldsymbol u_\beta \left[E(\boldsymbol u_\alpha,\boldsymbol u_\beta)\right]^t, 
\end{align}
where
\begin{align}
E(\boldsymbol u_\alpha,\boldsymbol u_\beta)=\expec{ \mathbb 1_2^{\otimes (\alpha-1)} \sigma_{\boldsymbol u_\alpha}\otimes \mathbb 1_2^{(\beta-\alpha-1)} \otimes \sigma_{\boldsymbol u_\beta} \otimes \mathbb 1_2^{\otimes (N-\beta)}}_\rho.
\label{app:TwoBodyCorrelationFct}
\end{align}
denotes the two-body correlation function resulting from measurements on qubits $\alpha$ and $\beta$, with $\alpha,\beta=1,\ldots,N$, and $\rho$ is an $N$-qubit density matrix. With the criterion $F(\mathcal R^{(2)}_{\alpha,\beta},\mathcal R^{(t)}_{\alpha,\beta})\geq 0$ we thus probe the entanglement of the corresponding two-body reduced state $\rho_{\alpha,\beta}=\text{tr}_{\{1,\ldots,N\}/\{\alpha,\beta\}}\left[\rho\right]$ of an $N$-qubit states $\rho$. To demonstrate multi-qubit entanglement detection based on two-body moments we use in the main text the class of Dicke states. As Dicke states are invariant under permutations of the involved subsystems, we can probe their entanglement by  applying $F(\mathcal R^{(2)}_{\alpha,\beta},\mathcal R^{(t)}_{\alpha,\beta})\geq 0$ to any of its $N$ qubits. In Fig.~2(b) of the main text we present the results of this procedure. To calculate the two-body moments of Dicke states for up to $200$ qubits and $100$ excitations we note that the two-body reduced density matrices of Dicke states $\ket{D^{N}_k}$ read \cite{DickeMarginals}:
\begin{align}
\text{tr}_{\{1,\ldots,N\}/\{\alpha,\beta\}}\left[\ket{D^{N}_k}\bra{D^{N}_k}\right]=&v_+ \ketbra{00}{00}+v_- \ketbra{11}{11}+y \ketbra{01}{01}+y \ketbra{01}{10}\\
&+y \ketbra{10}{01}+y \ketbra{10}{10},
\label{eq:DickeRedSts}
\end{align}
where
\begin{align}
v_+&=	\frac{(N-1)(N-k-1)}{N(N-1)},\\
v_-&=	\frac{k(k-1)}{N(N-1)},\\
y&=	\frac{k(N-k)}{N(N-1)}.
\end{align}
In the special case of $W$-states $\ket{W_N}=\ket{D_1^N}$ the two-body reduced states (\ref{eq:DickeRedSts}) read:
\begin{align}
\text{tr}_{\{1,\ldots,N\}\setminus\{i,j\}}\left(\ket{W_N}\bra{W_N}\right)=\frac{2}{N} \ket{\Psi^+}\bra{\Psi^+}+\frac{N-2}{N} \ket{00}\bra{00}
\end{align}
with the Bell state $\ket{\Psi^+}=\frac{1}{\sqrt 2}(\ket{01}+\ket{10})$.

\subsection{\label{app:E.2}E.2: With $N$-body moments}
Next we discuss entanglement detection based on moments of full $N$-body correlation functions as defined in Eq.~(\ref{app:RandomMoments}). First, we focus on the criteria involving only one of the random moments, namely $\mathcal R^{(2)}$, for which one obtains the criterion $\mathcal R^{(2)}\leq 1/3^N$, for all fully separable states. The derivation of this criterion was shortly reviewed at the end of Section~C and as it has been introduced and discussed previously (see \cite{tran1,tran2,BriegelLUinv}) we will not discuss it further at this point.

Further on, we investigate the entanglement detection capabilities when the moments $\mathcal R^{(2)}$ and $\mathcal R^{(4)}$ are combined. However, already for three qubits an analytical characterization of the set of separable states in the space of moments becomes very demanding and makes the use of numerical tools inevitable. For this reason, we develop numerical tools for generating states of different state subclasses. In the case of fully separable states this amounts to the generation of random pure product states which are subsequently mixed with random mixing parameters. This allows us to produce random fully separable states of different rank. The inset of Fig.~3(a) of the main text shows a plot of more than $10^5$ such fully separable states. 

For generation of mixed $N$-qubit states different strategies exist. A well known one is to draw random pure states from a higher dimensional Hilbert space and subsequently trace out over the extended dimensions. However, the latter approach leads to random states which are not homogeneously distributed in the space spanned by the moments $\mathcal R^{(2)}$ and $\mathcal R^{(4)}$. For this reason, we follow a different path motivated by the above strategy for generating fully separable states. Namely, we draw random pure states of the respective subclass and  mix them with randomly chosen mixing parameters. In particular, as the random moments are invariant under local unitary transformations, we can restrict ourselves to random pure states obtained from a so-called standard form. Standard forms are generalizations of the well-known two-qubit Schmidt decomposition to more than two qubits. A three qubit standard form was introduced in \cite{StForm3qAcin}:
\begin{align}
\ket{\Psi_{\text{s.f.}}}=\lambda_0 \ket{000}+\lambda_1e^{i\phi}\ket{100}+\lambda_2 \ket{101}+\lambda_3\ket{110}+\lambda_4\ket{111},
\label{app:3qubitStForm}
\end{align}
where $\lambda_i\leq0$, with $i=0,\ldots,4$, and $0\leq \phi\leq \pi$. A generalization Eq.~(\ref{app:3qubitStForm}) to $N$-qubits states has been introduced in \cite{StFormNq}. Standard forms allow also for an characterization of states of different SLOCC classes. For instance, with $\lambda_4=\phi=0$ and $\lambda_i>0$, for $i=0,1,2,3$, Eq.~(\ref{app:3qubitStForm}) becomes a standard form of the three-qubit $W$-class $\mathcal W^{(3)}$. A similar standard form for $\mathcal W^{(N)}$, with arbitrary $N$, was presented in Ref.~\cite{StForm3qTurgut}. In Fig.~3(a) of the main text we present the generated random states in the $(\mathcal R^{(2)},\mathcal R^{(4)})$-plane. In conclusion, we find that by combining the second and fourth moments a similar advantage for the detection of entanglement, as obtained for two-qubits in Fig.~1(a), is observed. Turning this numerical observation into an entanglement criteria is subject of future work. 

We end this section by explaining how the black lines in Fig.~3(a) of the main text are computed.  The black solid lines represent the border of the pure $W$-class. They have been calculated by maximizing the second and fourth moment over the standard form corresponding to the class $\mathcal W^{(3)}$. The black dashed lines indicate the additional area covered by states contained in the mixed $W$-class, namely $\text{Conv}(\mathcal W^{(3)})$. As the set $\text{Conv}(\mathcal W^{(3)})$ is more difficult to characterize, we estimated its boarder by minimizing the moments over the subset $\text{Conv}(\{\ketbra{000}{000},\ketbra{W_3}{W_3},\mathbb 1_3\})\subset \text{Conv}(\mathcal W^{(3)})$. We emphasize that this procedure leads only to an estimate of the borders of $\text{Conv}(\mathcal W^{(3)})$ which was confirmed by generating about $10^5$ random mixed $W$-states. This is in contrast to the black solid lines which are the result of a numerical optimization over the whole set $\mathcal W^{(3)}$. It is important to note, however, that if one aims at deriving linear criteria for the detection of states outside of $\text{Conv}(\mathcal W^{(N)})$, it is enough to resort to an optimization over $\mathcal W^{(N)}$, as discussed in the next Section.

\section{\label{app:F} F: Detection of states outside of the mixed $W$-class}

In this Section we outline the derivation of the criteria discussed in the last Section of the main text, which allow us to detect states that are not contained in the mixed $W$-class $\text{Conv}(\mathcal W^{(N)})$. There, we focus first on the second moment $\mathcal R^{(2)}$ and optimize it over the class of pure $W$-states $\mathcal W^{(N)}$. As $\mathcal R^{(2)}$ is invariant under local unitary transformation this optimization can be carried out through the respective $N$-qubit $W$-class standard form \cite{StForm3qTurgut}. Further on, the convexity of $\mathcal R^{(2)}$ leads to the following criterion: 
\begin{align}
\mathcal R^{(2)}_{\rho}\leq \sum_\alpha p_\alpha \mathcal R^{(2)}_{\rho'} \leq  \max\limits_{\rho'\in \mathcal W^{(N)}} \mathcal R^{(2)}_{\rho'}:=\chi^{(N)},
\label{app:GHZcritR2}
\end{align}
for all $\rho\in \text{Conv}(\mathcal W^{(N)})$. The maxima $\chi^{(N)}:=\max\limits_{\rho'\in \mathcal W^{(N)}} \mathcal R^{(2)}_{\rho'}$ can be evaluated analytically up to five qubits. To do so, we plug in the corresponding $W$-class standard form and optimize over the remaining parameters yielding $\chi^{(3)}=11/81$, $\chi^{(4)}=4/81$ and $\chi^{(5)}=7/405$, which are reached by the three-, four- and five-qubit $W$-state, respectively. Beyond five qubits we have to resort to numerical techniques to calculate the maxima $\chi^{(N)}$. The robustness of the criterion~(\ref{app:GHZcritR2}) is presented in Fig.~3(b) of the main text. There we plot the noise threshold $p^*$ up to which the noisy GHZ state $p \mathbb 1/2^N+(1-p)\ketbra{GHZ_N}{GHZ_N}$ can be detected by the witness~(\ref{app:GHZcritR2}). To check the performance of our criterion for pure states we apply it to  a GHZ-like state with variable amplitudes $\ket{\Psi(\theta)}=\cos{\theta} \ket{0}^{\otimes N}+\sin{\theta} \ket{1}^{\otimes N}$ and plot the threshold angle $\theta^*$ above which the state is detected to be outside of the $W$-class $\text{Conv}(\mathcal W^{(N)})$ y. 

The second criterion is constructed by combining the moments $\mathcal R^{(2)}$ and $\mathcal R^{(4)}$ linearly. For a given number $N$ of qubits we first maximize $\mathcal R^{(2)}$ and $\mathcal R^{(4)}$, respectively, over the $W$-class $\mathcal W^{(N)}$. We denote the arguments of these maxima as $\rho^{(1)}= \underset{{\rho'\in \mathcal W^{(N)}}}{\arg\max} \mathcal R^{(2)}$ and $\rho^{(2)}=\underset{{\rho'\in \mathcal W^{(N)}}}{\arg\max} \mathcal R^{(4)}$, respectively. Next, we calculate the slope of the line passing through the corresponding maximum points:
\begin{align}
 m^{(N)}=\frac{\mathcal R^{(4)}_{\rho^{(1)}}-\mathcal R^{(4)}_{\rho^{(2)}}}{\mathcal R^{(2)}_{\rho^{(1)}}-\mathcal R^{(2)}_{\rho^{(2)}}}.
 \end{align}
This yields the line equation $\mathcal R^{(4)}=m^{(N)} \mathcal R^{(2)}+b^{(N)}$, where 
\begin{align}
b^{(N)}=\frac{\mathcal R^{(4)}_{\rho^{(2)}}  \mathcal R^{(2)}_{\rho^{(1)}}-\mathcal R^{(4)}_{\rho^{(1)}}  \mathcal R^{(2)}_{\rho^{(2)}}}{\mathcal R^{(2)}_{\rho^{(1)}}-  \mathcal R^{(2)}_{\rho^{(2)}}}
\label{app:yintercept}
\end{align}
 denotes the corresponding $\mathcal R^{(4)}$-intercept. Finally, in order to obtain a GHZ-entanglement criterion, we have to maximize the $\mathcal R^{(4)}$-intercept $b^{(N)}$  over the $W$-class $\mathcal W^{(N)}$. This ensures that the defined line really touches the border of  $\mathcal W^{(N)}$ in the $(\mathcal R^{(2)},\mathcal R^{(4)})$-plane. Hence, we find that $m^{(N)} \mathcal R^{(2)}+\tilde b^{(N)}\geq \mathcal R^{(4)}$, for all $\rho\in \mathcal W^{(N)}$ and with $\tilde b^{(N)}=\max\limits_{\rho\in \mathcal W^{(N)}}  b^{(N)}$. Lastly, the convexity of the  moments yields:
\begin{align}
\mathcal R^{(4)}_\rho- m^{(N)} \mathcal R^{(2)}_\rho=\mathcal R^{(4)}_\rho+ m'^{(N)} \mathcal R^{(2)}_\rho\leq \sum_\alpha p_\alpha (\mathcal R^{(4)}_{\ketbra{W^{(N)}_\alpha}{W^{(N)}_\alpha}} +m'^{(N)} \mathcal R^{(2)}_{\ketbra{W^{(N)}_\alpha}{W^{(N)}_\alpha}})\leq \tilde b^{(N)},
\label{app:GHZcritR2R4}
\end{align}
for all $\rho\in \text{Conv}(\mathcal W^{(N)})$, and where $m'^{(N)}=-m^{(N)}>0$, for $N=3,4,5,6$. In Eq.~(\ref{app:GHZcritR2R4}) we used that a linear combination of two convex functions with positive coefficients yields again a convex function. This is possible because the slope $m^{(N)}$ is negative in the considered cases $N=3,4,5,6$. The performance of the obtained criterion (\ref{app:GHZcritR2R4}) is presented Fig.~3(b) of the main text in terms of the same states as before.


\begin{thebibliography}{99}
\bibliographystyle{unsrt}


\bibitem{HorodeckiReview} R. Horodecki, P. Horodecki, M. Horodecki, and K. Horodecki, Rev. Mod. Phys. \textbf{81}, 865 (2009).

\bibitem{NielsenChuang} M. A. Nielsen and I. L. Chuang, \textit{Quantum Computation and Quantum Information} (Cambridge University Press, New York, 2000).

\bibitem{flagship} A. Ac\'{\i}n, I. Bloch, H. Buhrman, T. Calarco, C. Eichler, J. Eisert, D. Esteve, N. Gisin, S. J. Glaser, F. Jelezko, S. Kuhr, M. Lewenstein, M. F. Riedel, P. O. Schmidt, R. Thew, A. Wallraff, I. Walmsley, F. K. Wilhelm, arXiv:1712.03773.

\bibitem{TelepReview} S. Pirandola, J. Eisert, C. Weedbrook, A. Furusawa, and S. L. Braunstein, Nat. Phys. \textbf{9}, 641 (2015).

\bibitem{CryptoReview} N. Gisin, G. Ribordy, W. Tittel, and H. Zbinden, Rev. Mod. Phys. \textbf{74}, 145 (2002).

\bibitem{OtfriedReview} O. G\"uhne and G. T\'oth, Phys. Rep. \textbf{474}, 1 (2009).

\bibitem{BELL}  J.S. Bell, Physics {\bf 1}, 195 (1964).

\bibitem{ReviewNonlocality} N. Brunner, D. Cavalcanti, S. Pironio, V. Scarani, and S. Wehner, Rev. Mod. Phys. \textbf{86}, 419 (2014).

\bibitem{GrossPolytopes} M. Walter, B. Doran, D. Gross, and M. Christandl, Science \textbf{340}, 1205 (2013).

\bibitem{StevePolytopes} G. H. Aguilar, S. P. Walborn, P. H. Souto Ribeiro, and L. C. C\'eleri, Phys. Rev. X \textbf{5}, 031042 (2015).

\bibitem{LeandroRotInv} L. Aolita and S. P. Walborn, Phys. Rev. Lett. \textbf{98}, 100501 (2007).

\bibitem{SteveAlignmentFreeComm} V. D'Ambrosio, E. Nagali, S. P. Walborn, L. Aolita, S. Slussarenko, L. Marrucci, and F. Sciarrino, Nat. Comm. \textbf{3}, 961 (2012).

\bibitem{BriegelLUinv} H. Aschauer, J. Calsamiglia,  M. Hein, and H. J. Briegel, Quantum Inf. Comput. \textbf 4, 383 (2004).

\bibitem{JulioCorrMat1} J. I. de Vicente, Quantum Inf. Comput. \textbf 7, 624 (2007).

\bibitem{JulioCorrMat2} J. I. de Vicente, J. Phys. A: Math. Theor. \textbf{41}, 065309 (2008).

\bibitem{JulioMarkus} J. I. de Vicente and M. Huber, Phys. Rev. A \textbf{84}, 062306 (2011).

\bibitem{ZukowskiRefFrame1} P. Badziag, C. Brukner, W. Laskowski, T. Paterek, and M. \.Zukowski, Phys. Rev. Lett. \textbf{100}, 140403 (2008).

\bibitem{ZukowskiRefFrame2} W. Laskowski, M. Markiewicz, T. Paterek, and M. \.Zukowski, Phys. Rev. A \textbf{84}, 062305 (2011).

\bibitem{LawsonRefFrame} T. Lawson, A. Pappa, B. Bourdoncle, I. Kerenidis, D. Markham, and E. Diamanti, Phys. Rev. A \textbf{90}, 042336 (2014).

\bibitem{KlocklHuber} C. Kl\"ockl and M. Huber, Phys. Rev. A \textbf{91}, 042339 (2015).

\bibitem{ReviewRefFrames} S. D. Bartlett, T. Rudolph, and R. W. Spekkens, Rev. Mod. Phys. \textbf{79}, 555 (2007).

\bibitem{tran1} M. C. Tran, B. Daki\'c, F. Arnault, W. Laskowski, and T. Paterek, Phys. Rev. A \textbf{92}, 050301(R) (2015).

\bibitem{tran2} M. C. Tran, B. Daki\'c, W. Laskowski, and T. Paterek, Phys. Rev. A \textbf{94}, 042302 (2016).

\bibitem{LukRud} A. Gabriel, \L. Rudnicki, and B. C. Hiesmayr, New J. Phys. \textbf{15}, 073033 (2013).

\bibitem{RudolphRandMeasBell} Y.-C. Liang, N. Harrigan, S. D. Bartlett, and T. Rudolph, Phys. Rev. Lett. \textbf{104}, 050401 (2010).

\bibitem{BrunnerRandMeasBell} P. Shadbolt, T. V\'ertesi, Y.-C. Liang, C. Branciard, N. Brunner, and J. L. O'Brien, Sci. Rep. \textbf 2, 470 (2012).

\bibitem{DiamantiRandMeasBell} C. Furkan Senel, T. Lawson, M. Kaplan, D. Markham, and E. Diamanti, Phys. Rev. A \textbf{91}, 052118 (2015).


\bibitem{existence} P. D. Seymour, and T. Zaslavsky, Advances in Mathematics \textbf{52}, 213 (1984).

\bibitem{Dankert} C. Dankert, M.Sc. thesis, University of Waterloo, (2005); also available as e-print quant-ph/0512217.

\bibitem{SpinSqueez1} G. Toth, C. Knapp, O. G\"uhne, and H. J. Briegel, Phys. Rev. Lett. \textbf{99}, 250405 (2007).

\bibitem{SpinSqueez2} D. J. Wineland, J. J. Bollinger, W. M. Itano, and D. J. Heinzen, Phys. Rev. A \textbf{50}, 67 (1994).

\bibitem{SpinSqueez3} A. S\o{}rensen and K. M\o{}lmer, Phys. Rev. Lett. \textbf{83}, 2274 (1999).

\bibitem{SpinSqueez4} A. S\o{}rensen, L.-M. Duan, J. I. Cirac, and P. Zoller, Nature (London) \textbf{409}, 63 (2001).


\bibitem{ApproxDesign1} F. G. S. L. Brand\~ao, A. W. Harrow, and M. Horodecki, Phys. Rev. Lett. \textbf{116}, 170502 (2016).

\bibitem{ApproxDesign2} Y. Nakata, C. Hirche, M. Koashi, and A. Winter, Phys. Rev. X \textbf 7, 021006 (2017).

\bibitem{ExamplesSphericalDesigns} R. H. Hardin and N. J. A. Sloane, Discrete \& Computational Geometry \textbf{15}, 429 (1996).

\bibitem{Gross5Design} D. Gross, K. Audenaert, and J. Eisert, J. Math. Phys. \textbf{48}, 052104 (2007).

\bibitem{BellDiagStates} R. Horodecki and M. Horodecki, Phys. Rev. A \textbf{54}, 1838 (1996).

\bibitem{DickeMarginals} X. Zheng-Jun, X. Heng-Na, L. Yong-Ming, and W. Xiao-Guang, Commun. Theor. Phys. \textbf{57}, 771 (2012).

\bibitem{SLOCC3qDuer} W. D\"ur, G. Vidal, and J. I. Cirac, Phys. Rev. A \textbf{62}, 062314 (2000).

\bibitem{SLOCC3qAcin} A. Ac\'{\i}n, D. Bru\ss, M. Lewenstein, and A. Sanpera, Phys. Rev. Lett. \textbf{87}, 040401 (2001).

\bibitem{Conny4qu} C. Spee, J. I. de Vicente, and B. Kraus, J. Math. Phys. \textbf{57}, 052201 (2016).

\bibitem{StForm3qAcin} A. Ac\'{\i}n, A. Andrianov, L. Costa, E. Jan\'e, J. I. Latorre, and R. Tarrach, Phys. Rev. Lett. \textbf{85}, 1560 (2000).

\bibitem{StFormNq} H. A. Carteret, A. Higuchi, and A. Sudbery, J. Math. Phys. \textbf{41}, 7932 (2000).

\bibitem{StForm3qTurgut} S. K\i nta\c s and S. Turgut, J. Math. Phys. \textbf{51}, 092202 (2010).

\bibitem{Cliff3Design} Z. Webb, Quantum Inf. Comput. \textbf{16}, 1379 (2016).

\bibitem{CliffNo4Design} H. Zhu, R. Kueng, M. Grassl, and D. Gross, arXiv:1609.08172.

\bibitem{FreeSpaceQC} R. Ursin, et al., Nat. Phys. \textbf 3, 481 (2007).

\bibitem{SatelliteQC1} J. G. Rarity, P. R. Tapster, P. M. Gorman, and P. Knight, New J. Phys. \textbf 4, 82 (2002).

\bibitem{SatelliteQC2} M. Aspelmeyer, et al., Science \textbf{301}, 621 (2003).

\bibitem{SatelliteQC3} P. Villoresi, et al., New J. Phys. \textbf{10}, 033038 (2008).

\bibitem{SatelliteQC4} C. Bonato, A. Tomaello, V. Da Deppo, G. Naletto, and P. Villoresi, New J. Phys. \textbf{11}, 045017 (2009).

\bibitem{RandUExp} M. Bourennane, M. Eibl, S. Gaertner, C. Kurtsiefer, A. Cabello, and H. Weinfurter, Phys. Rev. Lett. \textbf{92}, 107901 (2004).

\end{thebibliography}
\end{document}